\documentclass[a4paper]{jpconf}
\usepackage{lineno,graphicx}
\DeclareUnicodeCharacter{2212}{\ensuremath{-}}
\begin{document}
\title{Hadronic resonance production with ALICE at the LHC}

\author{Sergey Kiselev$^*$ for the ALICE Collaboration}
%\author{Sergey Kiselev$^*$}

%\address{$^*$NRC "Kurchatov Institute" - ITEP, B. Cheremushkinskaya 25, 117218 Moscow, Russia}
%\address{$^*$NRC $<<$Kurchatov institute$>>$, Moscow, Russia}
\address{$^*$NRC ``Kurchatov Institute'', Moscow, Russia}

\ead{Sergey.Kiselev@cern.ch}

%\linenumbers
\begin{abstract}
Recent results on short-lived hadronic resonances obtained with the ALICE detector at LHC energies are presented. 
These results include masses, widths, transverse momentum spectra, 
yields, and ratios of resonance yields to longer-lived ground-state particles, and elliptic flows. 
The results are compared with model predictions and measurements at lower energies.
% and measurements at lower energies.
\end{abstract}

The study of resonance production is important in proton-proton, proton-nucleus, and heavy-ion collisions.
Since the lifetimes of short-lived resonances are comparable with the lifetime of the late hadronic phase 
produced in heavy-ion collisions, resonance yields are affected by the regeneration and rescattering 
of their decay daughters in the hadronic phase. These competing effects are investigated by measuring the yield ratios 
of resonances to that of the ground-state longer-lived hadron as a function of charged-particle multiplicity. 
From these measurements, it is possible to obtain information on the time interval between the chemical and the kinetic freeze-out.
Measurements in pp and p--Pb collisions constitute a reference for nuclear collisions 
and provide information for tuning event generators inspired by Quantum Chromodynamics.
Moreover, some heavy-ion effects (elliptic flow~\cite{ALICEflow}-\cite{ALICEflow2}, strangeness enhancement~\cite{ALICEstrenh})
 were also unexpectedly observed in small collision systems.

Results on short-lived 
mesonic $\rho(770)^{0}$, $\mathrm{K}^{*}(892)^{0}$, $\mathrm{K}^{*}(892)^{\pm}$, $f_{0}(980)$, $\phi(1020)$, $f_{1}(1275)$ 
as well as baryonic $\Sigma(1385)^{\pm}$, $\Lambda(1520)$ $\Xi(1530)^{0}$ and $\Omega(2012)^{-}$ resonances 
(hereafter denoted as $\rho^{0}$, $\mathrm{K}^{*0}$, $\mathrm{K}^{*\pm}$, $f_{0}$, $\phi$, $f_{1}$, $\Sigma^{*\pm}$, $\Lambda^{*}$, $\Xi^{*0}$, $\Omega^{*}$) 
have been obtained using data reconstructed with the ALICE detector.
The  resonances  are  reconstructed  in  their  hadronic  decay  channels  and have very different lifetimes
as shown in Table~\ref{tab:Res}. 
\begin{table}[ht]
\caption{Lifetime values~\cite{PDG}, reconstructed decay mode, reactions and the corresponding references where ALICE results for the hadronic resonances are reported.}
% and ALICE papers for hadronic resonances.}
\begin{center}
%\begin{tabular}{|c|c|c|c|c|c|}
\begin{tabular}{ c c c l l }
%\hline
\br
    Resonance      & \it{c}$\tau$(fm) &     Decay     &        System @ energy(TeV)      & ALICE papers \\
%                    &$\rho^{0}$&$\mathrm{K}^{*0}$  &$\mathrm{K}^{*\pm}$  & $f_{0}$ &  $\Sigma^{*\pm}$ &    $\Lambda^{*}$    &     $\Xi^{*0}$   &  $\phi$    \\
%\hline
\mr
%decay channel&$\pi\pi $&$\mathrm{K}\pi $&$\mathrm{K^{0}_{S}}\pi $&$\pi\pi $&$\Lambda\pi $&$p\mathrm{K} $&$\Xi\pi $&$\mathrm{K}\mathrm{K} $\\
$\rho^{0}$         &  1.3  & $\pi\pi$ & pp/p-Pb @ 2.76 & \cite{ALICErho0} \\
$\mathrm{K}^{*0}$  &  4.3  & $\mathrm{K}\pi$ & pp/p–Pb/Pb–Pb/Xe–Xe @ all energies & \cite{ALICEpp7}-\cite{ALICEppXe} \\ 
$\mathrm{K}^{*\pm}$&  4.3  & $\mathrm{K^{0}_{S}}\pi$ & pp @ 5.02/8/13  Pb–Pb @ 5.02 & \cite{ALICEppKstar-pm}-\cite{ALICEpp13Kstar-pm} \\
$f_{0}$            & $\sim$ 5  & $\pi\pi$ & pp/p–Pb @ 5.02 & \cite{ALICEppf0}-\cite{ALICEpp13f0} \\
$\Sigma^{*\pm}$    & 5-5.5 & $\Lambda\pi$ & pp @ 7/13 p–Pb/Pb–Pb @ 5.02 &\cite{ALICEppSigmaStar}-\cite{ALICEPbPbSigmaStar} \\
$f_{1}$            &  8.7  & $\mathrm{K^{0}_{S}}\mathrm{K}\pi$ & pp @ 13 & \cite{ALICEppf1} \\
$\Lambda^{*}$      & 12.6  & $p\mathrm{K}$ & pp @ 7 p–Pb @ 5.02 Pb–Pb@ 5.02 & \cite{ALICELambdaStar}-\cite{ALICELambdaStar2} \\
$\Xi^{*0}$         & 21.7  & $\Xi\pi$ & pp @ 7/13 p–Pb @ 5.02 & \cite{ALICEppSigmaStar}-\cite{ALICEpp13SstarXstar} \\
$\Omega^{*}$       & 32.3  & $\Xi\mathrm{K^{0}_{S}}$ & pp @ 13 & \cite{ALICEppOmegaStar} \\
$\phi$             & 46.2  & $\mathrm{K}\mathrm{K}$ & pp/p–Pb/Pb–Pb/Xe–Xe @ all energies & \cite{ALICEpp7}-\cite{ALICEpPb502},\cite{ALICEphiPb502Flow}-\cite{ALICEphiXe} \\
%\hline
%lifetime (fm/\it{c})& 1.3 & 4.2 & 4.2 & $\sim$ 5& 5-5.5 & 12.6 & 21.7& 46.2 \\ 
%ALICE papers &\cite{ALICErho0}&\cite{ALICEpp7}-\cite{ALICEppXe} &\cite{ALICEppKstar-pm} -\cite{ALICEPb5Kstar-pm} &\cite{ALICEppf0}-\cite{ALICEpPbf0} &\cite{ALICEppSigmaStar}-\cite{ALICEPbPbSigmaStar} &\cite{ALICELambdaStar}-\cite{ALICELambdaStar2} &\cite{ALICEppSigmaStar}-\cite{ALICEpp13SstarXstar} &\cite{ALICEpp7}-\cite{ALICEpPb502},\cite{ALICEphiXe}-\cite{ALICEphiPbFlow} \\
\br
\end{tabular}
\end{center}
 \label{tab:Res}
\end{table}
%

%The V0A and V0C detectors (32 scintillating counters each) were used for the determination of 
%the multiplicity classes by measuring the sum of the signals from V0A and V0C forming the V0M signal.
 
This contribution reports recent results obtained
for $f_{1}$  \cite{ALICEppf1}, $\Omega^{*}$ \cite{ALICEppOmegaStar}, $\mathrm{K}^{*\pm}$ \cite{ALICEpp13Kstar-pm} and $f_{0}$
\cite{ALICEpp13f0} in pp collisions at 13 TeV, for $\mathrm{K}^{*0}$ and $\phi$ in Pb--Pb collisions at 5.36 TeV,  and for $f_0(1710)$ in pp collisions at 13.6 TeV.

\begin{figure}[hbtp]
\begin{center}
\includegraphics[scale=0.37]{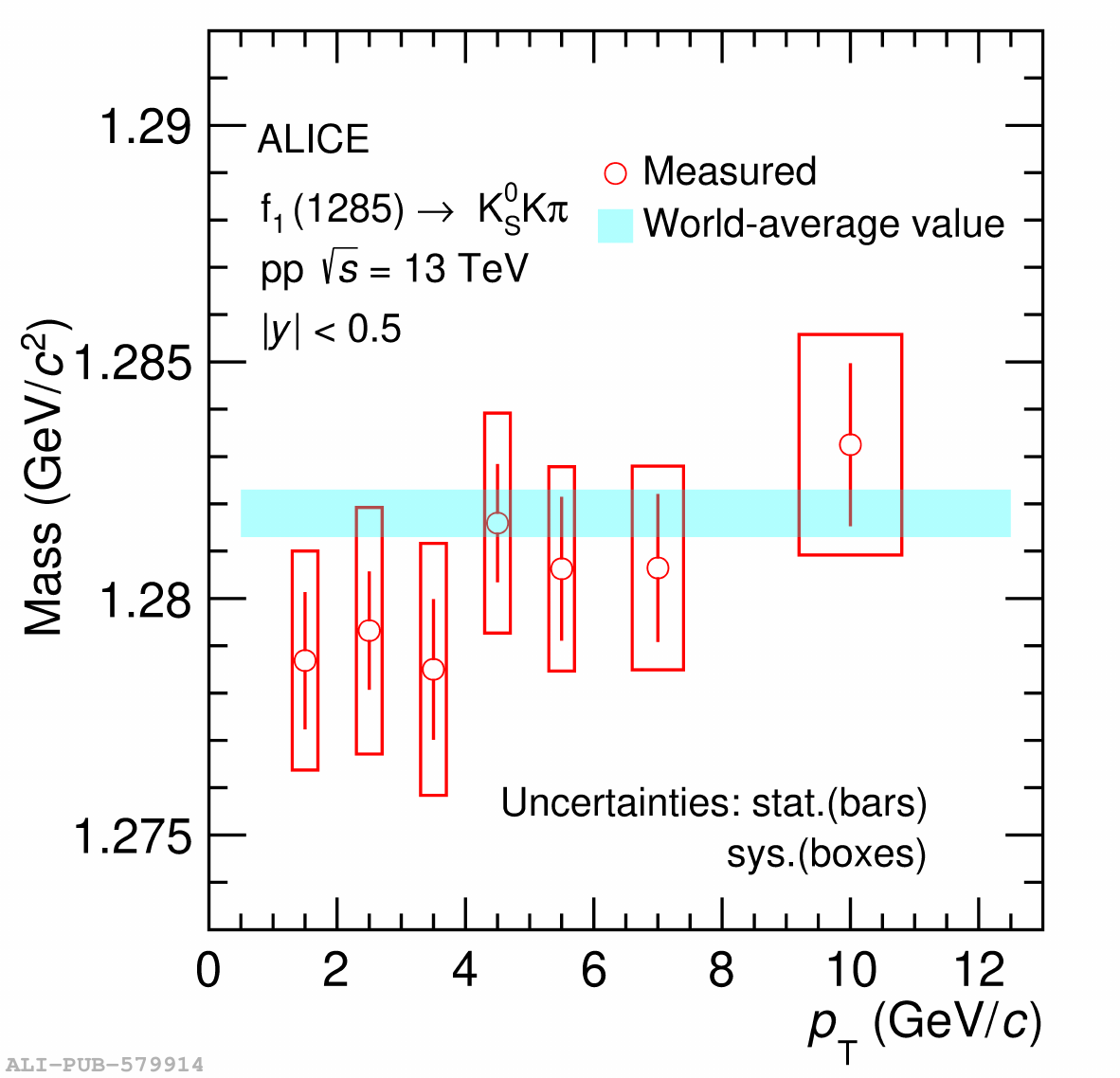}
\includegraphics[scale=0.37]{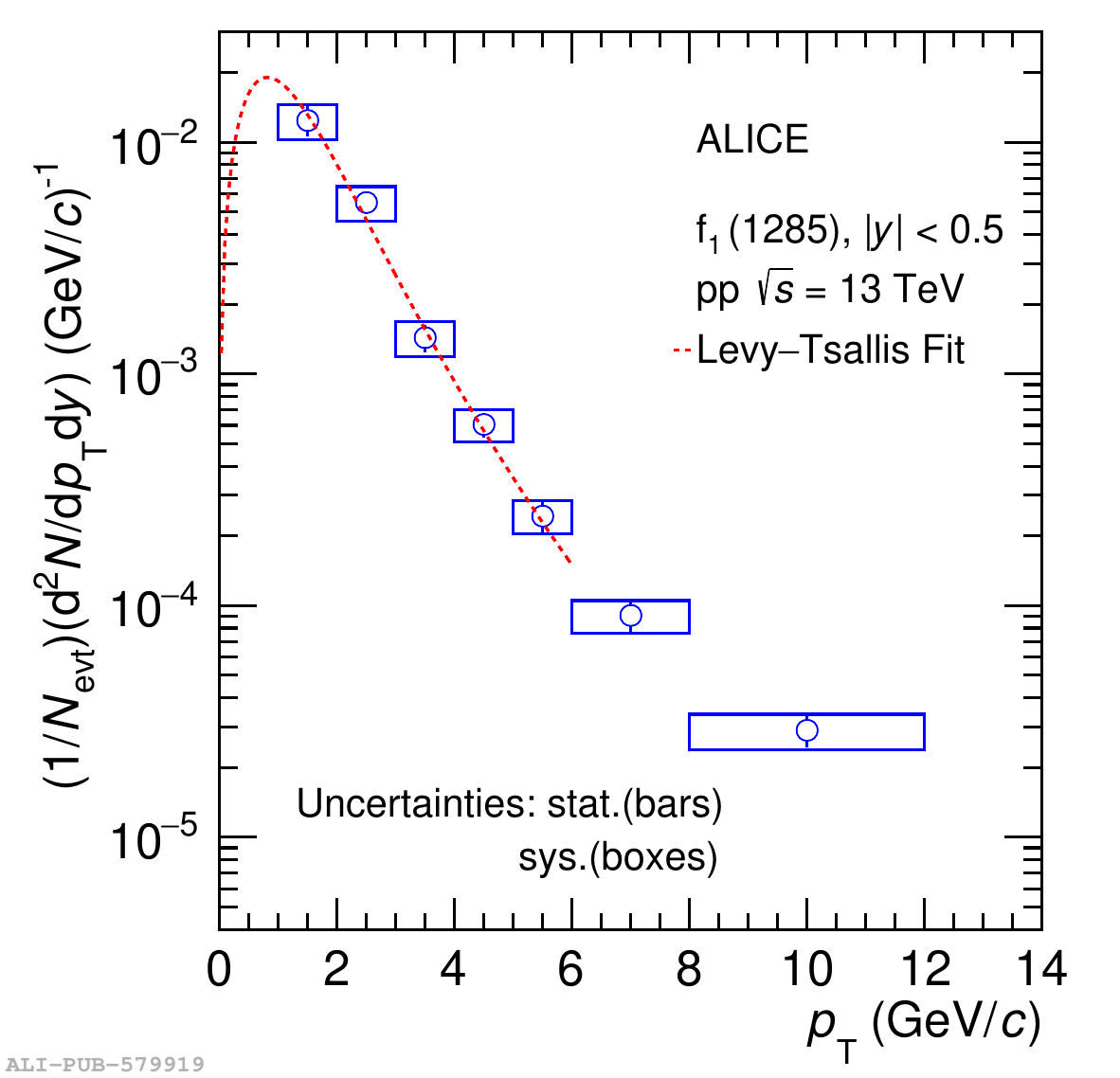}
\includegraphics[scale=0.37]{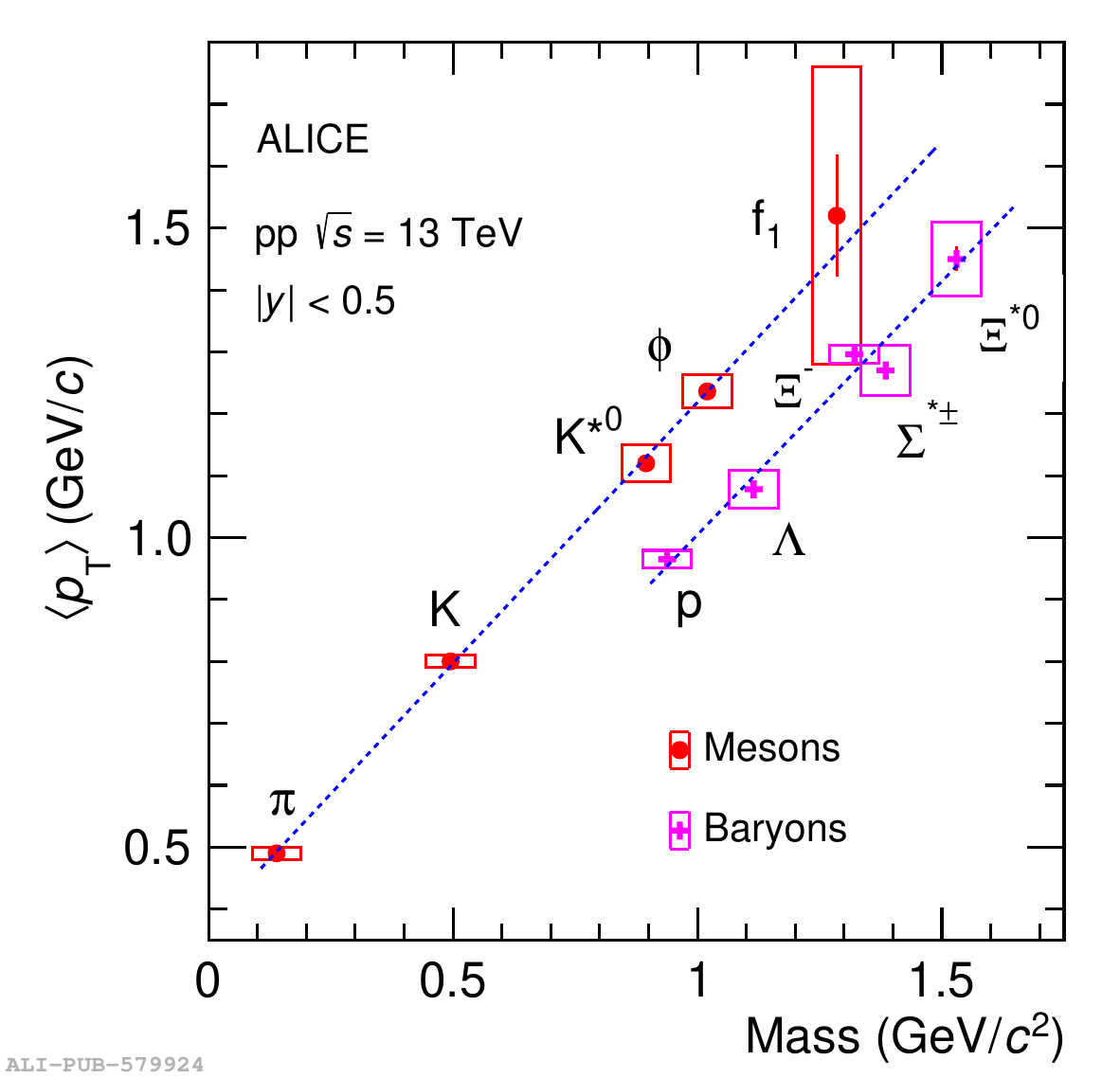}
\includegraphics[scale=0.37]{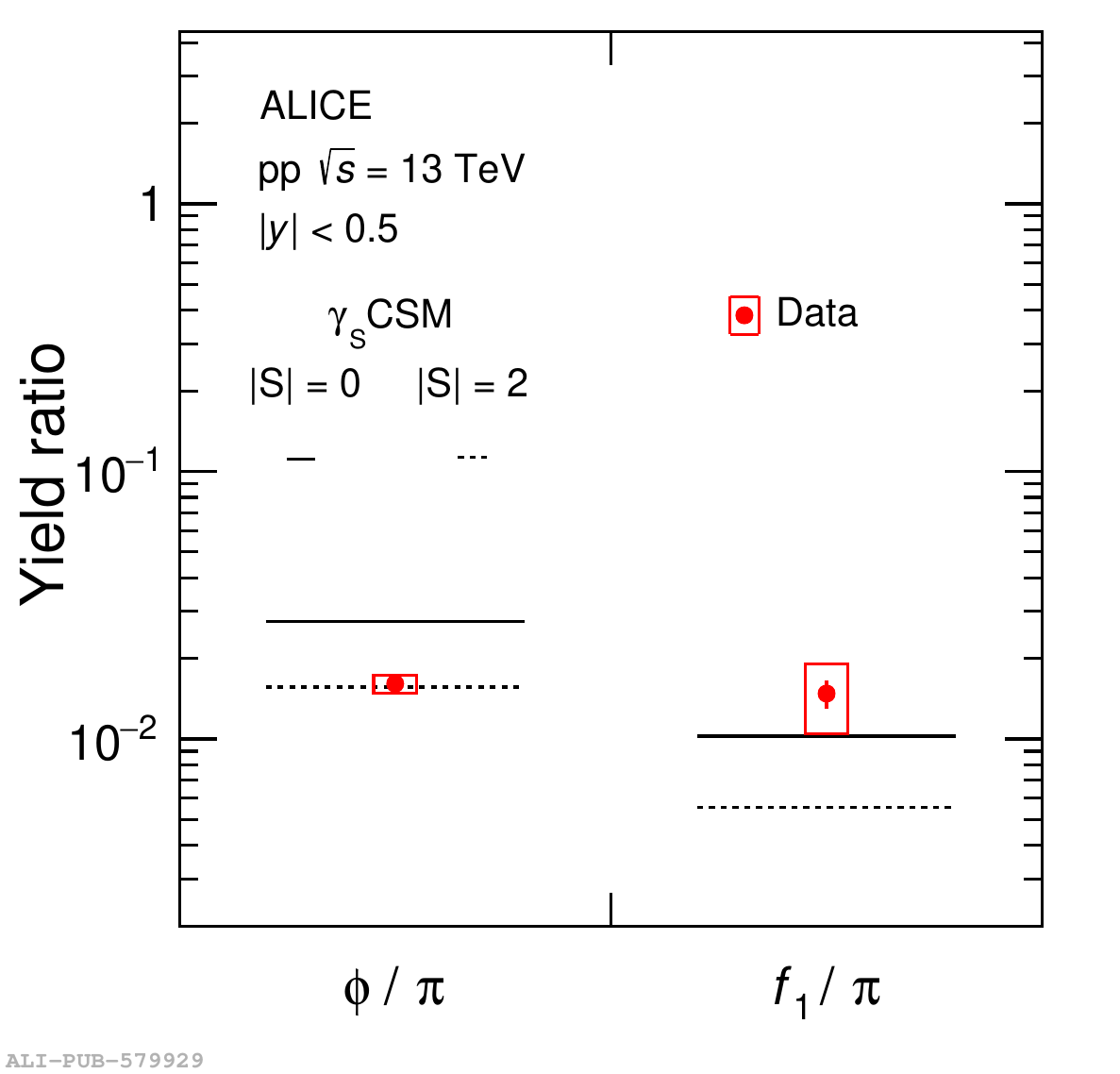}
\end{center}
\caption{ 
Mass (top, left) and  $p_\mathrm{T}$-differential yield (top, right) of the $f_{1}$.
The $p_\mathrm{T}$ spectrum is fitted by a Levy–Tsallis \cite{Levy-Tsallis} function. 
(bottom, left) The average transverse momentum of $f_{1}$ with that of all other 
light flavor hadrons \cite{ALICEpp713}, \cite{ALICEpp13SstarXstar}.
(bottom, right) The particle yield ratios $\phi/\pi$ \cite{ALICEpp713} and $f_{1}/\pi$ with model predictions 
from  $\gamma_{S}$-CSM \cite{GsCSM}.
}
  \label{fig:MassSpec}
\end{figure}

%The ALICE Collaboration made 
The first measurement of the $f_{1}$ meson has been made in pp collisions at $\sqrt{s}$ = 13 TeV. 
The mass of $f_{1}$ is in reasonable agreement with the world-average value, 
~Fig.~\ref{fig:MassSpec} (top, left). 
Figure~\ref{fig:MassSpec}  (top, right) shows the $p_\mathrm{T}$ spectrum with
a fit by a Lévy–Tsallis function, to extrapolate the yield down to zero $p_\mathrm{T}$ and extract
the $p_\mathrm{T}$-integrated yield d$N$/d$y$ and the average transverse momentum $\langle p_\mathrm{T}\rangle$.
Since there are only two $p_\mathrm{T}$ bins above 6 GeV/$c$ with large bin width,
the Lévy–Tsallis fit in the default case is performed in the $0 < p_\mathrm{T} < 6$ GeV/$c$ range.
The extrap­olation to the low $p_\mathrm{T}$ ( $<$ 1 GeV/$c$) region encompasses approximately
41\% of the total $f_{1}$ yield. The high $p_\mathrm{T}$ extrapolation is found to be negligible.
Figure~\ref{fig:MassSpec} (bottom, left) compares the average transverse momentum of $f_{1}$ with that of all other 
light flavor hadrons. For particles with similar masses, mesons exhibit a higher average transverse momentum 
than baryons. Notably, $f_{1}$ aligns with the linear trend of other mesons, although with
large conservative systematic uncertainties. 
This observation suggests that $f_{1}$ may have an ordinary meson structure.
Figure~\ref{fig:MassSpec} (bottom, right) shows the particle yield ratios $\phi/\pi$ and $f_{1}/\pi$ with model predictions from  $\gamma_{S}$-CSM for $|S|$=0 (zero strangeness content) 
and $|S|$=2 (hidden strangeness content). As expected, for $\phi/\pi$ the ratio with $|S|$=2 is in good 
agree­ment with the data. For $f_{1}/\pi$, the ratio with $|S|$=0 agrees with the data within $1\sigma$,
and with $|S|$=2 deviates by $\sim 2\sigma$, suggesting that $f_{1}$ is a conventional non-strange meson.

\begin{figure}[hbtp]
\begin{center}
\includegraphics[scale=0.37]{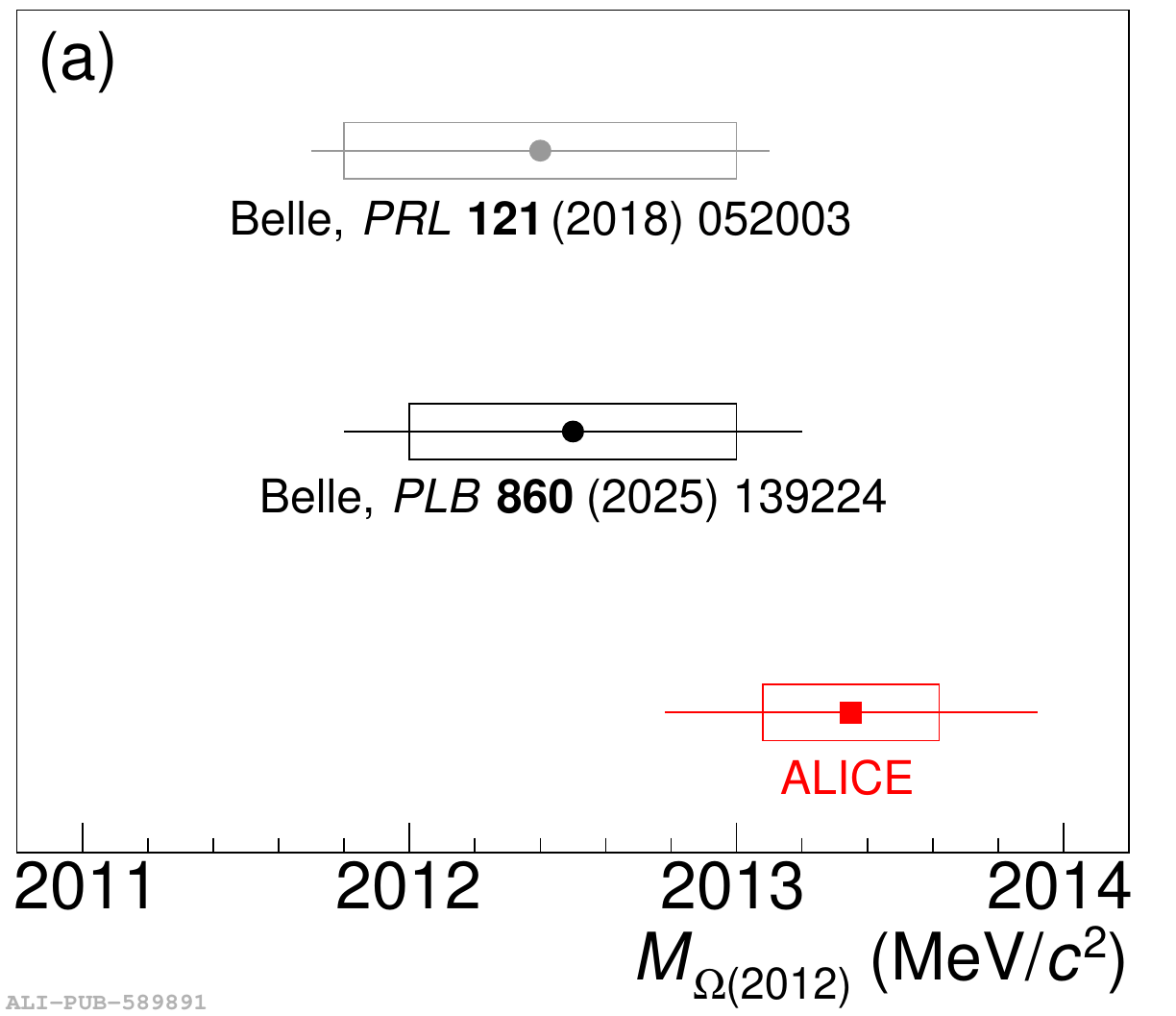}
\includegraphics[scale=0.37]{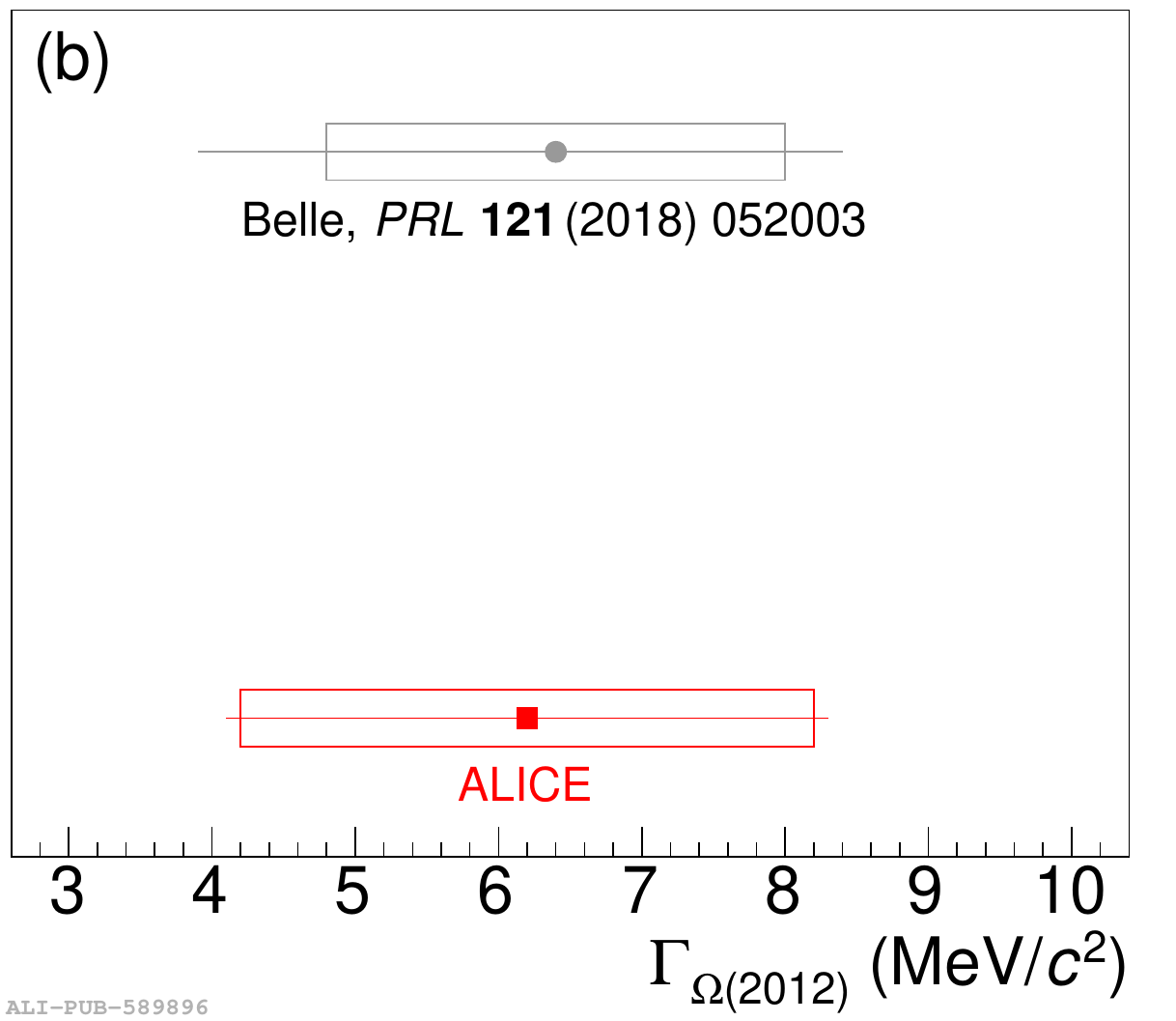}
\end{center}
\caption{
Mass (a) and width (b) of the $\Omega^{*}$. 
Results of Belle \cite{Belle1}, \cite{Belle2} are also shown. 
}
  \label{fig:mpt}
\end{figure}

\begin{figure}[hbtp]
\begin{center}
\includegraphics[scale=0.5]{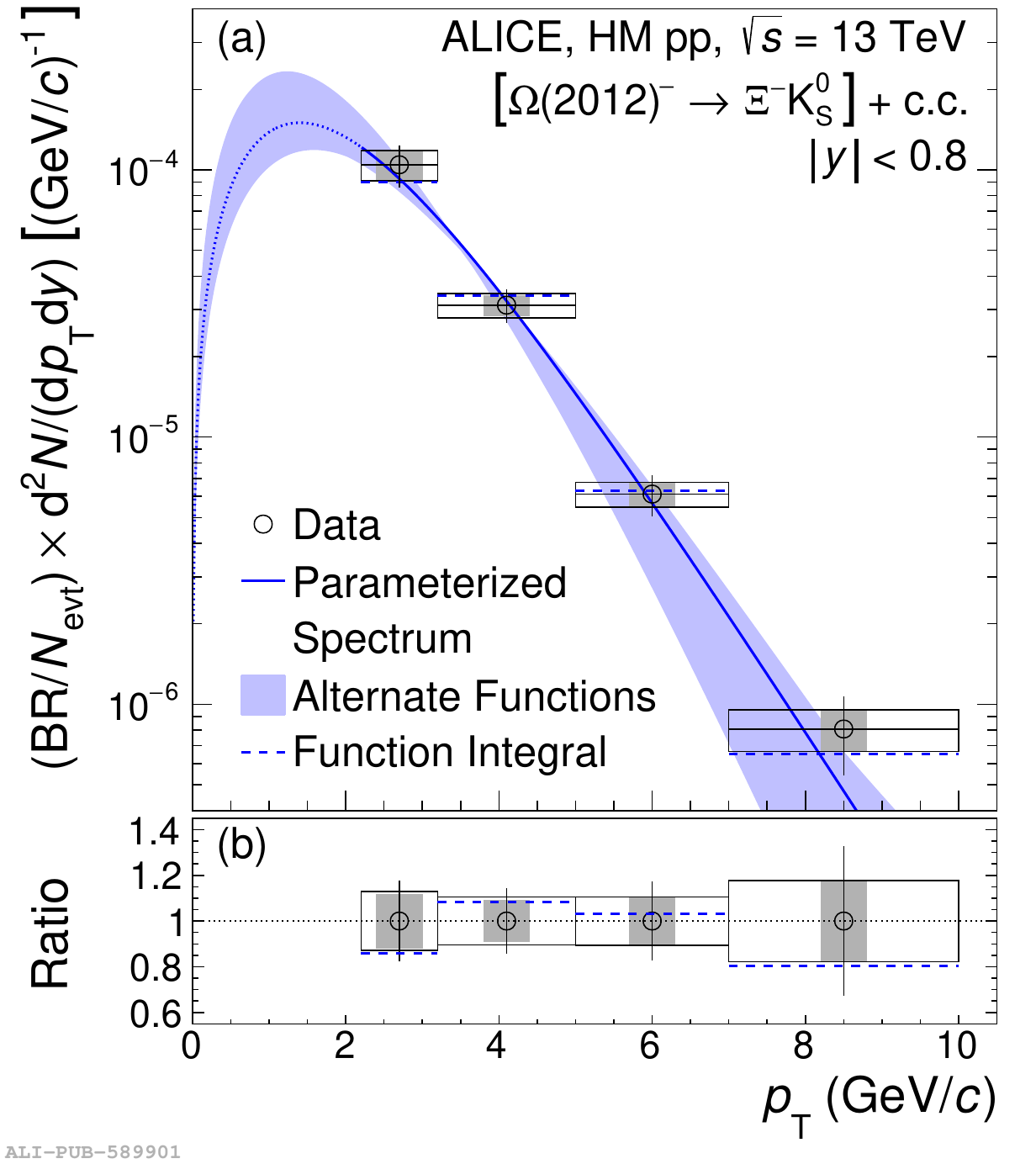}
\end{center}
\caption{
(a) The transverse-momentum spectrum of the $\Omega^{*}$ in HM pp collisions at $\sqrt{s}$ = 13 TeV (not corrected for
the unmeasured branching ratio BR for the studied decay channel).
The shaded band surrounding the curve indicates the region spanned by the envelope of alternate functions used to describe the spectrum, which affect the extrapolation of the yield to low $p_\mathrm{T}$. 
(b) The ratio of the parameterized yield to the measured yield.
}
  \label{fig:Mratios}
\end{figure}

A signal consistent with the $\Omega^{*}$ baryon has been observed with a significance of 15$\sigma$ 
in high-multiplicity-triggered (HM) pp collisions at $\sqrt{s}$ = 13 TeV.
%The $\Omega^{*}$ was reconstructed via the $\Omega^{*} \rightarrow \Xi^{-} \mathrm{K^{0}_{S}}$ decay channel and its charge conjugate.
Figure~\ref{fig:mpt} shows the mass and width of $\Omega^{*}$. 
These values are consistent with the previous measurements by Belle.
The first measurement of transverse-momentum spectrum has been made.
The $\Omega^{*}$ transverse-momentum spectrum, not corrected for the unmeasured 
$\Omega^{*} \rightarrow \Xi \mathrm{K^{0}_{S}}$ branching ratio, is shown in Fig.~\ref{fig:Mratios}.
The spectrum is fitted with the parameterized spectrum function, with only the overall scale factor allowed to vary.
%Based on a comparison to the statistical thermal model expectation, the total branching ratio for the Ω(2012) − → ΞK decays is estimated to be 0.62 +0.27 −0.17 .

\begin{figure}[hbtp]
\begin{center}
\includegraphics[scale=0.44]{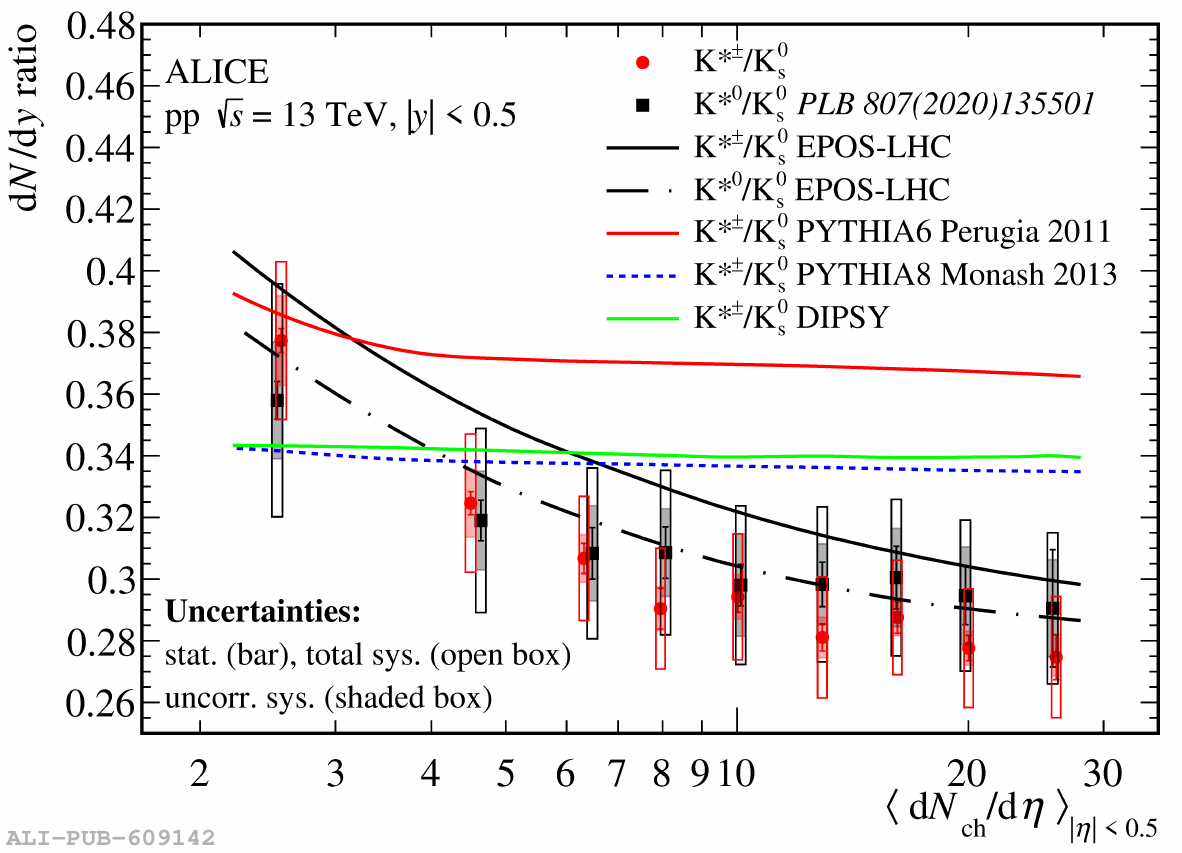}
\includegraphics[scale=0.33]{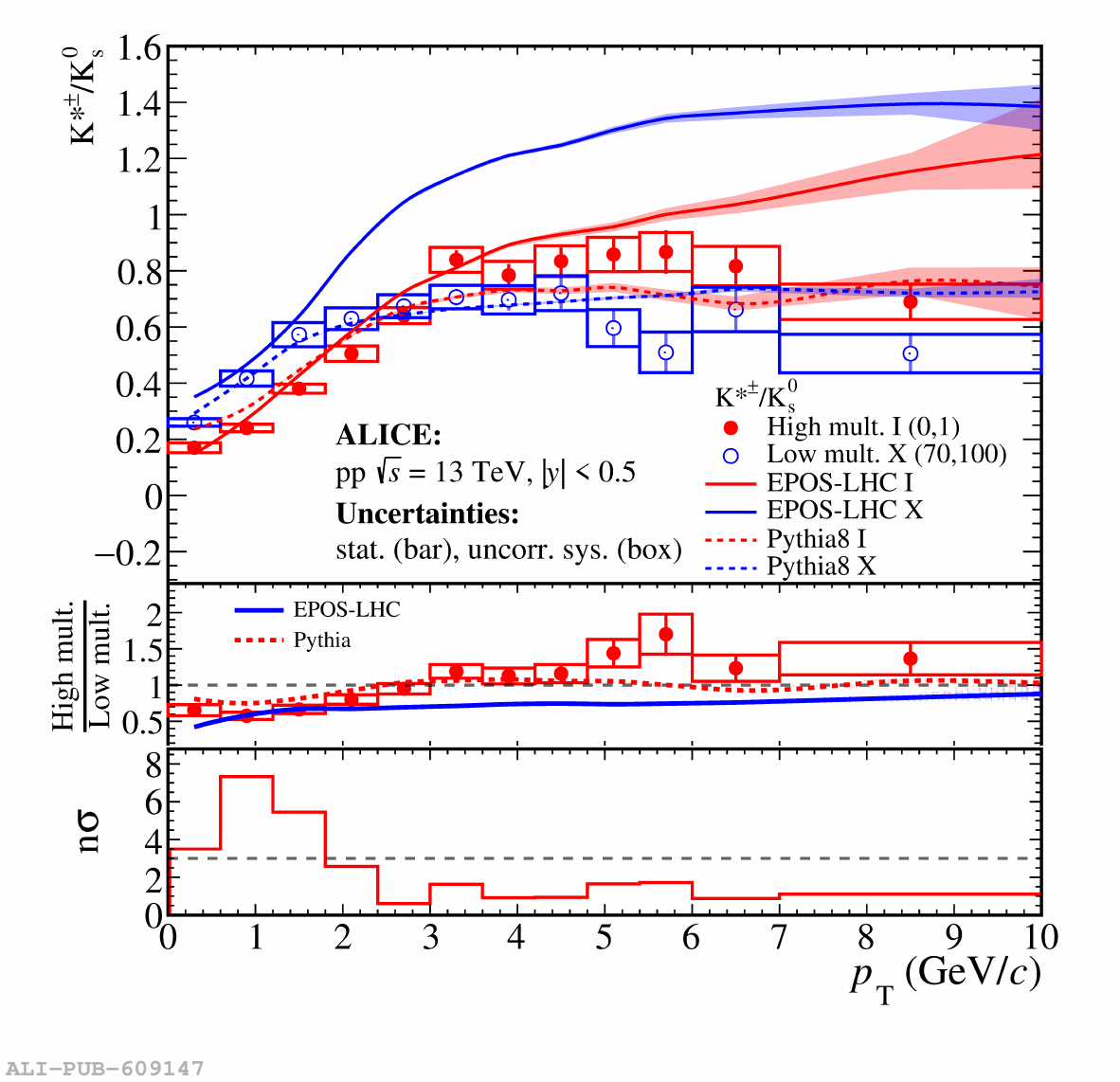}
\end{center}
\caption{(color online)
Left: Ratios $\mathrm{K}^{*\pm}/\mathrm{K}$ and $\mathrm{K}^{*0}/\mathrm{K}$~\cite{ALICEpp13} as a function of the charged-particle multiplicity density
in pp collisions at $\sqrt{s}$ = 13 TeV.
Model predictions from PYTHIA6 \cite{Perugia}, PYTHIA8 \cite{Monash}, EPOS-LHC \cite{EPOSLHC} and DIPSY \cite{DIPSY}
%, $\gamma_{S}$-CSM (gammaS-CSM) \cite{GsCSM} and HRG \cite{HRG-PCE} 
are also shown. 
Right: Ratios $\mathrm{K}^{*\pm}/\mathrm{K}$ as a function of $p_\mathrm{T}$ for low (X) and high (I) multiplicity classes. 
}
  \label{fig:Kratio13}
\end{figure}

Figure~\ref{fig:Kratio13} (left) shows the multiplicity dependence of the $\mathrm{K}^{*\pm}/\mathrm{K}$ ratio in pp collisions at $\sqrt{s}$ = 13 TeV, 
compared to the $\mathrm{K}^{*0}/\mathrm{K}$ ratio.
The decreasing trend already outlined by the $\mathrm{K}^{*0}$ analysis is confirmed by the $\mathrm{K}^{*\pm}$ results.
The $\mathrm{K}^{*\pm}/\mathrm{K}$ ratio in the highest multiplicity class is lower than the low multiplicity value at $\sim$ 7$\sigma$ level 
taking into account the multiplicity uncorrelated uncertainties ($\sim$ 2$\sigma$ level for the $\mathrm{K}^{*0}/\mathrm{K}$ ratio).
This result represents the first evidence of a clear $\mathrm{K}^{*}/\mathrm{K}$ suppression measured in small collision systems. EPOS-LHC
% \cite{EPOSLHC} 
provides good agreement with the measured data, well reproducing the decreasing trend, while PYTHIA6
%  \cite{Perugia}
, PYTHIA8
%   \cite{Monash}
and DIPSY
% \cite{DIPSY} 
tend to overestimate the ratios at high multiplicities and exhibit a fairly flat trend.
It is worth noting that none of these event generators consider the evolution of
a hadronic phase or the eventual hadronic interactions of generated particles.
The ratio of the high multiplicity $\mathrm{K}^{*\pm}/\mathrm{K}$ differential $p_\mathrm{T}$ distribution to the low multiplicity one helps 
to quantify the observed decrease in the particle ratios, Fig.~\ref{fig:Kratio13} (right). For $p_\mathrm{T} \leq$ 2 GeV/$c$, the  double $\mathrm{K}^{*\pm}/\mathrm{K}$ ratio 
deviates from unity by more than 3$\sigma$, suggesting a low $p_\mathrm{T}$ dominant process.

\begin{figure}[hbtp]
\begin{center}
\includegraphics[scale=0.37]{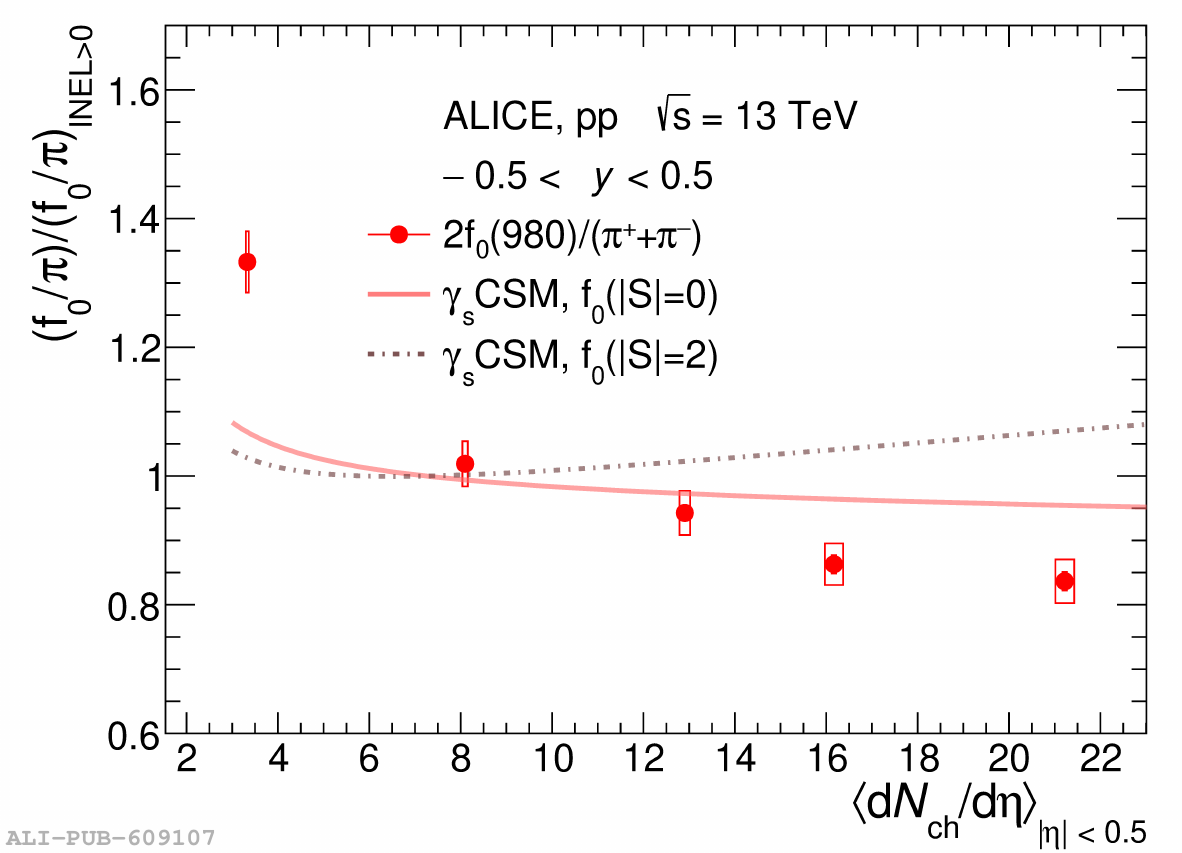}
\includegraphics[scale=0.37]{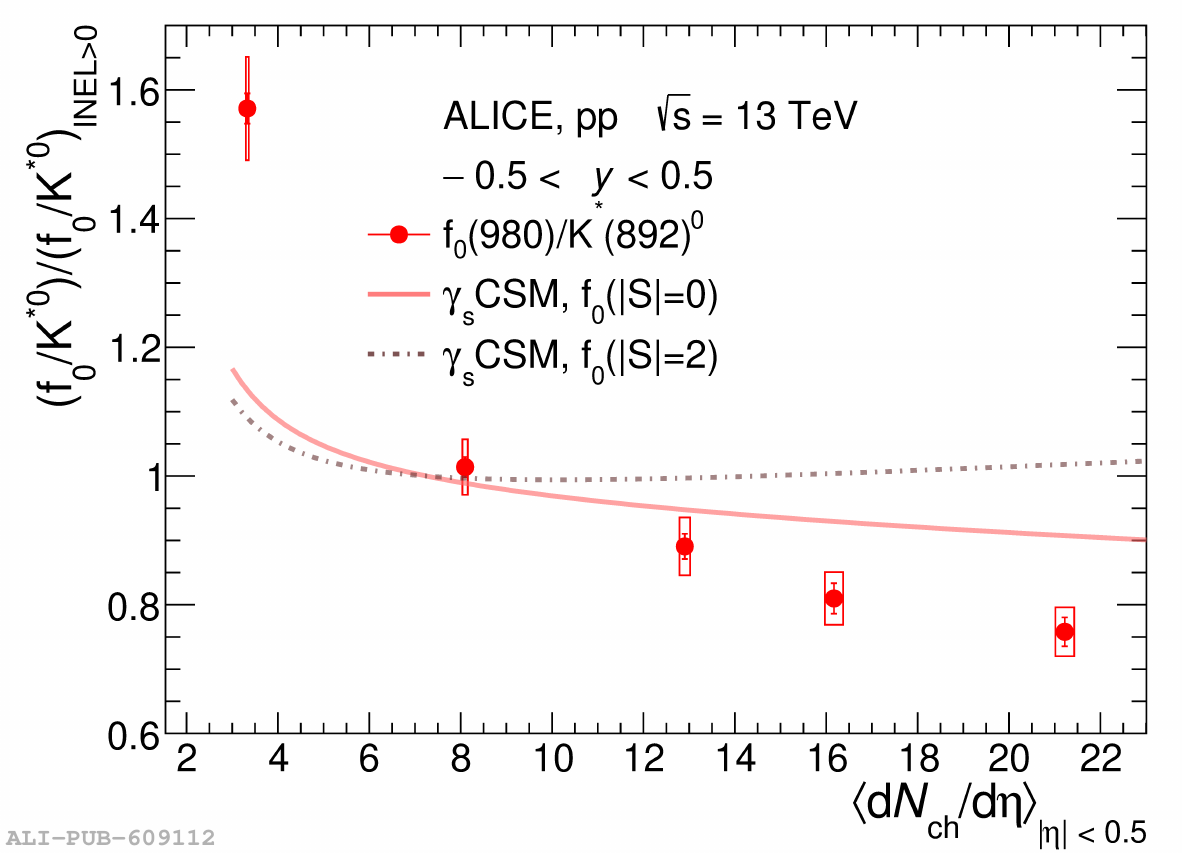}
\end{center}
\caption{
Double ratios of $f_{0}$ to $\pi$ (left) and $\mathrm{K}^{*0}$ (right) as a function of multiplicity in pp collisions at $\sqrt{s_\mathrm{NN}}$  = 13 TeV.
Model predictions from $\gamma_{S}$-CSM \cite{GsCSM} are also shown.
%Right: $p_\mathrm{T}$-differential particle yield ratio of $f_{0}$ to pion. 
}
  \label{fig:f0}
\end{figure}

Figure~\ref{fig:f0} demonstrates  the double ratios of $f_{0}$ to $\pi$ (left) and $\mathrm{K}^{*0}$ (right) as a function of multiplicity in pp collisions at $\sqrt{s}$  = 13 TeV.
The $f_{0}/\pi$ suppression could be explained by the dominance of the rescattering effect.
The double ratio $f_{0}/\mathrm{K}^{*0}$ also decreases with increasing multiplicity.
Even if rescattering affects both $f_{0}$ and $\mathrm{K}^{*0}$, these resonances, which have similar lifetimes, differ by their decay daughters: ($\pi$,$\pi$) for $f_{0}$ and (K,$\pi$) for $\mathrm{K}^{*0}$. The rescattering effect for $f_{0}$ can be stronger than for $\mathrm{K}^{*0}$ if the cross section $\sigma(\pi \pi) > \sigma(K \pi$).
The $\gamma_{S}$-CSM (where rescattering is not implemented) predictions with $|S|$=0 scenario are closer to the data values. However, the model does not reproduce the ratios at low multiplicity, in contrast to other hadrons \cite{GsCSM}.
% suggesting that $f_{0}$ is a conventional meson.
%good description of particle yield ratio in low multiplicity events

\begin{figure}[hbtp]
\begin{center}
\includegraphics[scale=0.5]{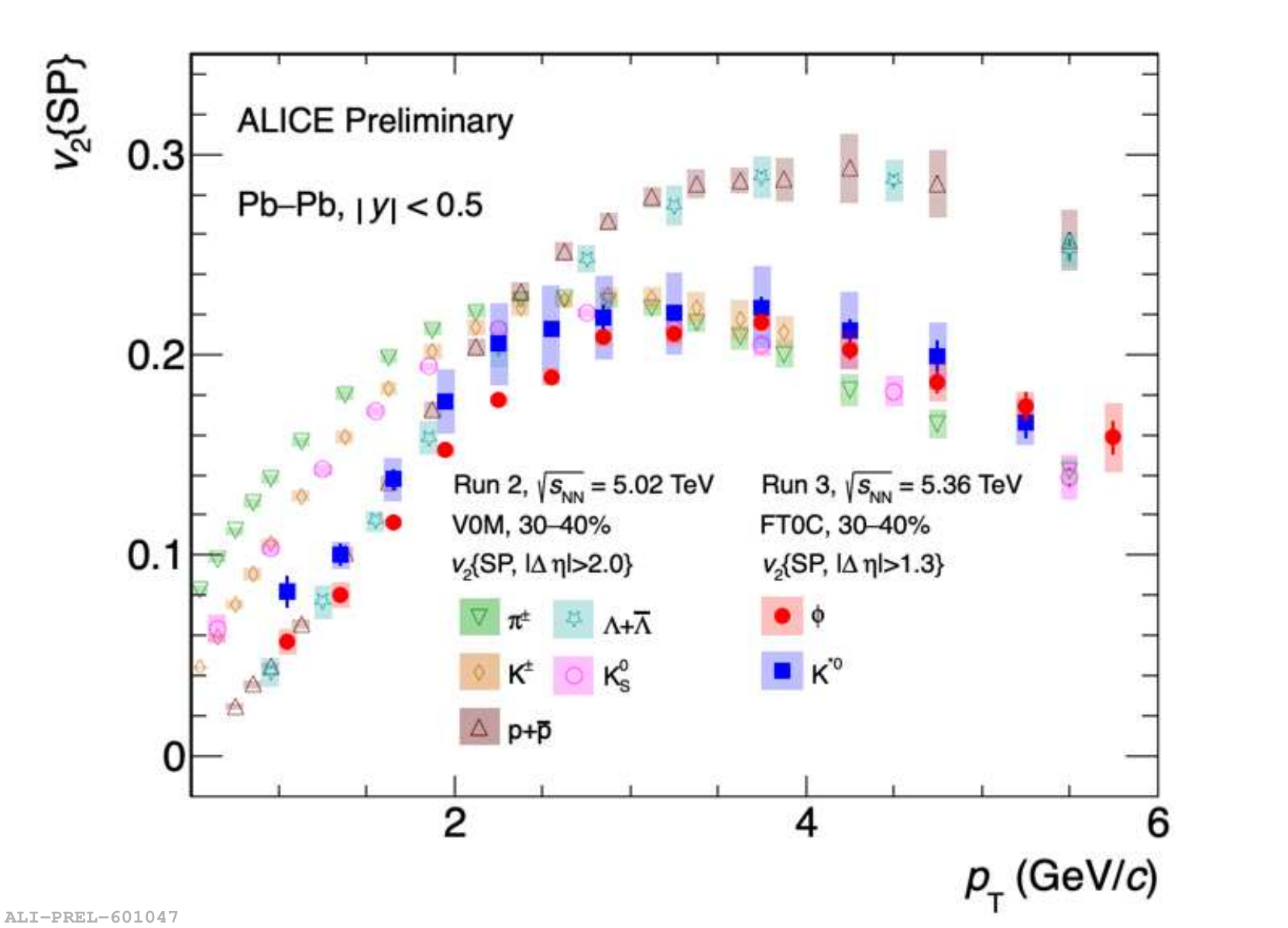}
\end{center}
\caption{
The elliptic flow for $\phi$ and $\mathrm{K}^{*0}$ in Pb--Pb collisions at $\sqrt{s_{\mathrm {NN}}}$ = 5.36 TeV from Run 3.
Measurements for other hadrons from Run 2 \cite{ALICEphiPb502Flow} are also shown.
%Comparison of the $v_2$ of $\phi$ and $\mathrm{K}^{*0}$ from Run 3 with measurements of
%other hadrons from Run 2 for Pb--Pb collisions with centrality 30–40%.
}
  \label{fig:flow}
\end{figure}

Figure~\ref{fig:flow} shows the $\phi$ and $\mathrm{K}^{*0}$ meson elliptic flow as a function of $p_\mathrm{T}$ in Pb–Pb collisions at $\sqrt{s_{\mathrm {NN}}}$ = 5.36 TeV, along with the ALICE measurements for other hadrons in Pb–Pb collisions at $\sqrt{s_{\mathrm {NN}}}$ = 5.02 TeV for centrality class 30–40\%. 
%QGP.
%
Both $\mathrm{K}^{*0}$ and $\phi$ follow mass ordering at low $p_\mathrm{T} \leq$ 3 GeV/$c$ and baryon-meson grouping 
at intermediate $p_\mathrm{T}$ . These observations confirm that both the $\mathrm{K}^{*0}$ and $\phi$ mesons participate in the collective expansion of the medium and undergo hadronization via quark coalescence in 
quark–gluon plasma.

\begin{figure}[hbtp]
\begin{center}
\includegraphics[scale=0.35]{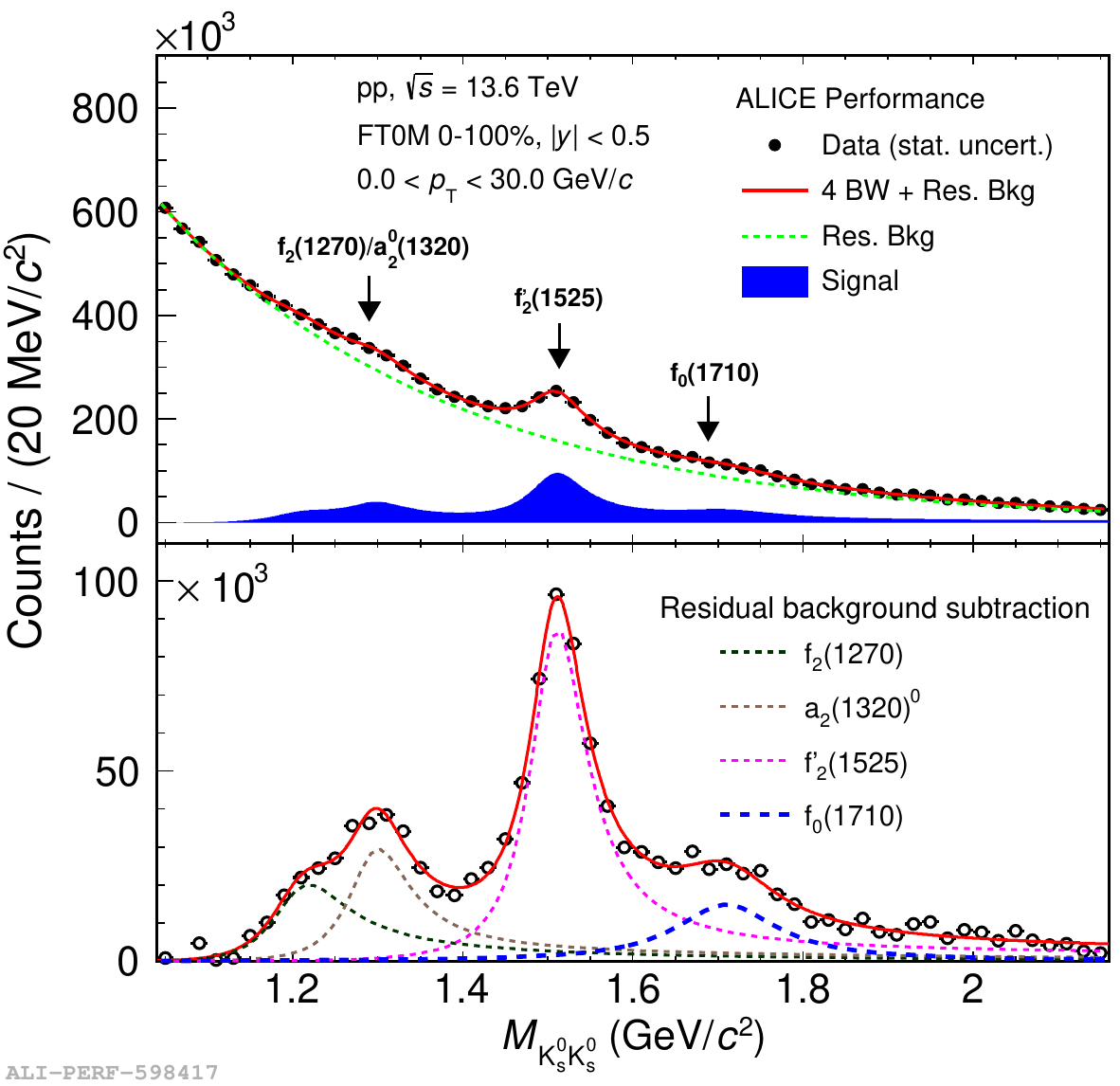}
\includegraphics[scale=0.35]{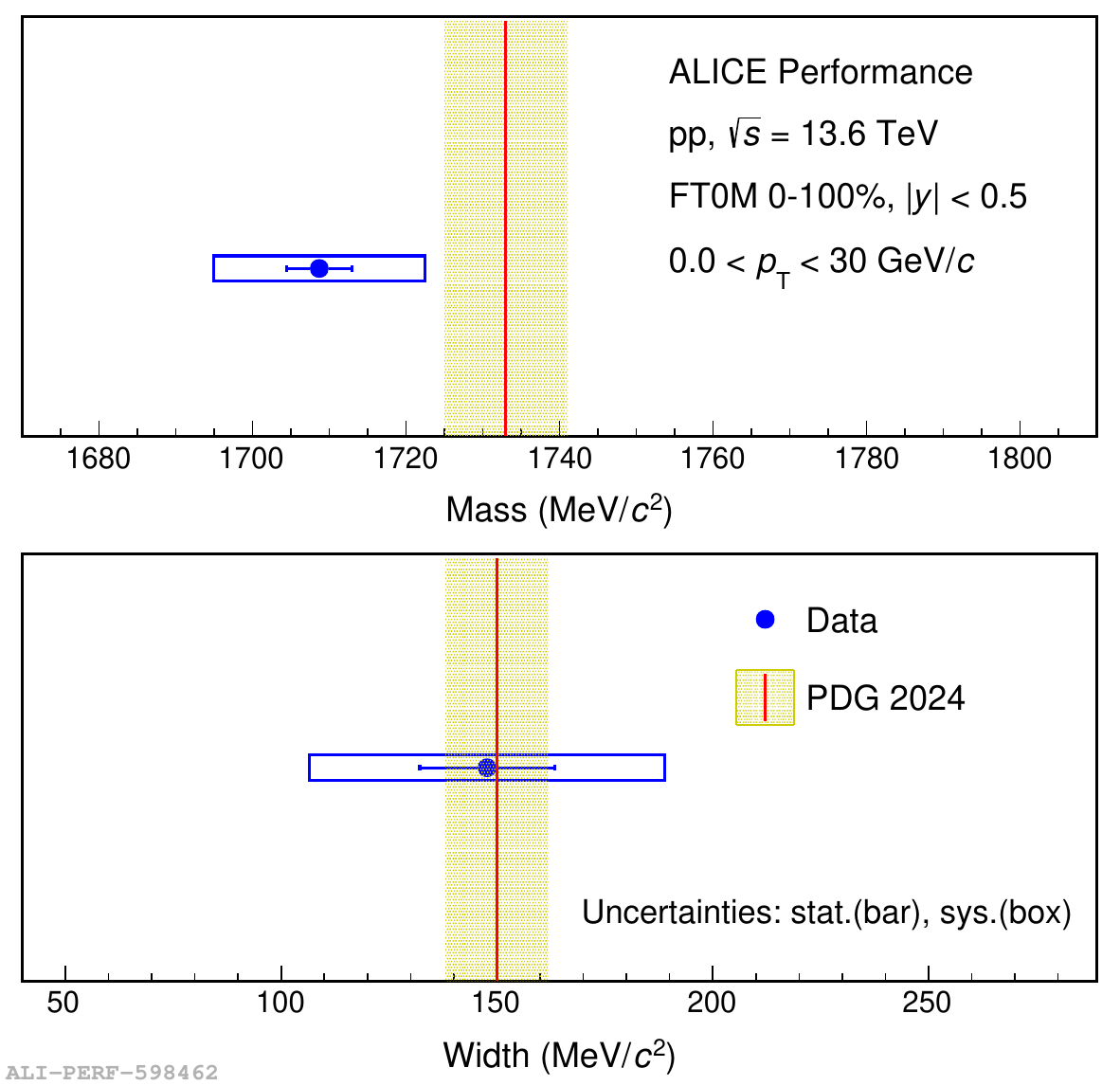}
\end{center}
\caption{
Left: Background subtracted $\mathrm{K^{0}_{S}} \mathrm{K^{0}_{S}}$ invariant mass distribution in pp collisions 
at $\sqrt{s}$ = 13.6 TeV. Right: Mass and width of $f_0(1710)$.
}
  \label{fig:f01770}
\end{figure}

Figure~\ref{fig:f01770} (left) presents the $\mathrm{K^{0}_{S}} \mathrm{K^{0}_{S}}$ invariant mass distribution in pp collisions 
at $\sqrt{s}$ = 13.6 TeV.
The $f_0(1710)$ meson is observed. It can be the lightest scalar glueball candidate predicted by QCD \cite{Glueballs}.
%Mass and width (Fig.~\ref{fig:f01770}, right) are close to the PDG values.
The mass of $f_0(1710)$ is within $1.5\sigma$ of the PDG value, the width aligns with the PDG value (Fig.~\ref{fig:f01770}, right).  

In summary, recent results on short-lived hadronic resonances $f_{1}$, $\Omega^{*}$, $\mathrm{K}^{*\pm}$ and $f_{0}$
in pp collisions at $\sqrt{s}$ = 13 TeV, $\mathrm{K}^{*0}$, $\phi$ in Pb--Pb at $\sqrt{s_{\mathrm {NN}}}$ = 5.36 TeV, and $f_0(1710)$ in pp at 13.6 TeV have been presented.
% Fig.1c
The average transverse momentum of $f_{1}$ aligns with the linear trend with mass observed for other mesons.
This observation suggests that $f_{1}$ may have an ordinary meson structure.
% Fig.1d
The $\gamma_{S}$-CSM model prediction with $|S|$=0 (zero strangeness content) for the $f_{1}/\pi$ ratio agrees with the data within $1\sigma$ and with $|S|$=2 (hidden strangeness content) deviates by $\sim 2\sigma$ suggesting that $f_{1}$ is a conventional meson.
% Figs.2-3
The mass and width of $\Omega^{*}$ are consistent with the previous measurements by Belle.
Firstly the transverse-momentum spectrum of the $\Omega^{*}$ has been measured.
% Fig.4
The $\mathrm{K}^{*\pm}/\mathrm{K}$ ratio in the highest multiplicity class is lower than the low multiplicity value 
at $\sim$ 7$\sigma$ level representing the first evidence of a $\mathrm{K}^{*}/\mathrm{K}$ suppression measured 
in pp collisions. The EPOS-LHC model without any hadronic afterburner is able to reproduce the measured suppression.
% Fig.5
Both the $f_{0}/\pi$ and $f_{0}/\mathrm{K}^{*0}$ ratios decrease with increasing multiplicity.
The $f_{0}/\pi$ suppression could be explained by the dominance of the rescattering effect.
The rescattering effect for $f_{0}$ can be stronger than for $\mathrm{K}^{*0}$ if cross section $\sigma(\pi \pi) > \sigma(K \pi$).
For both resonances $\gamma_{S}$-CSM (no rescattering effect) predictions with $|S|$=0 scenario are closer to the data values.
% Fig.6
The elliptic flow for $\phi$ and $\mathrm{K}^{*0}$
follow mass ordering at low $p_\mathrm{T} \leq$ 3 GeV/$c$ and baryon-meson grouping 
at intermediate $p_\mathrm{T}$ . These observations confirm that both the $\mathrm{K}^{*0}$ and $\phi$ mesons participate in the collective expansion of the medium and undergo hadronization via quark coalescence in 
quark–gluon plasma.
% Fig.7
%Mass and width of the $f_0(1710)$ meson observed in pp collisions at $\sqrt{s}$ = 13.6 TeV are close to the PDG values.
The mass of $f_0(1710)$ is within $1.5\sigma$ of the PDG value, the width aligns with the PDG value.  

\section*{Acknowledgement}
The work was carried out within the state assignment of NRC ``Kurchatov Institute''.

\end{document}